\documentclass[11pt]{article}
\usepackage{moriond,epsfig}
\usepackage{url}

\newcommand{\squishlist}{
 \begin{list}{$\bullet$}
  { \setlength{\itemsep}{1pt}
     \setlength{\parsep}{3pt}
     \setlength{\topsep}{3pt}
     \setlength{\partopsep}{1pt}
     \setlength{\leftmargin}{2em}
     \setlength{\labelwidth}{1em}
     \setlength{\labelsep}{0.5em} } }
\newcommand{\squishend}{
  \end{list}  }
\def\be{\begin{equation}}
\def\ee{\end{equation}}
\def\bea{\begin{eqnarray}}
\def\eea{\end{eqnarray}}

\begin{document}
\vspace*{4cm}
\title{Preliminary results of charged pions cross-section in proton
  carbon interaction at 30 GeV measured with the NA61/SHINE detector.}

\author{Sebastien Murphy \\on behalf of the NA61/SHINE collaboration \url{http://na61.web.cern.ch/}}

\address{University of Geneva, DPNC, 24 Quai E. Ansermet, 1201 Geneva, Switzerland}

\maketitle\abstracts{
%Long baseline neutrino osciilltation experiment are limited by 
  As the intensity of neutrino beams produced at accelerators
  increases, important systematic errors due to poor knowledge of
  production cross sections for pions and kaons arise.  Among other
  goals, the NA61/SHINE (SHINE $\equiv$ SPS Heavy Ion and Neutrino
  Experiment) detector at CERN SPS aims at precision hadro-production
  measurements to characterise the neutrino beam of the T2K experiment
  at J-PARC.  These measurements are performed using a 30 GeV proton
  beam produced at the SPS with a thin carbon target and a full T2K
  replica target. Preliminary spectra of $\pi^{-}$ and $\pi^{+}$
  inclusive cross section were obtained from pilot data collected in
  2007 with a 2 cm thick target. After a description of the SHINE
  detector and its particle identification capabilities, results from
  three different analysis are discussed.}

\section{Physics motivation}
In T2K, neutrinos are produced by a high intensity proton beam of 30
GeV impinging on a carbon target and producing mesons ($\pi$ and $K$)
from the decay of which the neutrinos are produced. There exist so far
no measurements of hadron inclusive spectra from p+C at 30 GeV.
Thus the NA61/SHINE experiment will provide a precise
measurement of meson yield production in carbon at the
proton beam energy (30 GeV/c) of interest for T2K. These
measurements will be used for the T2K neutrino beam simulation
and consequently reduce the systematic uncertainties of the neutrino
energy distribution at the needed level for the physics goals
of T2K \cite{nico}.

\section{The SHINE detector and combined particle identification}
\begin{figure}[h!]
	\centering
        \includegraphics[width=.7\textwidth,height=.25\textheight]{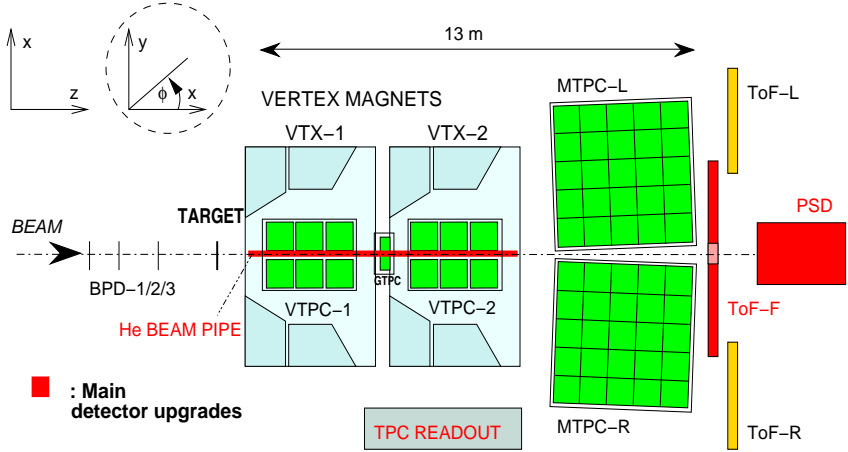}
	\caption{\label{fig:layout}The layout of the NA61/SHINE set-up
          in the 2007 data taking (top view, not to scale).}
\end{figure}

The set-up of the NA61/SHINE is shown in Fig.~\ref{fig:layout}. The
main components of the NA61 detector were inherited from the NA49
experiment \cite{NA49}. The tracking apparatus consits in four large
volume Time Projection Chambers (TPCs), which are capable of detecting
up to 70\% of all charged particles created in the reactions
studied. Two vertex TPCs (VTPC-1 and VTPC-2), are located in the
magnetic field of two super-conducting dipole magnets and, two TPCs
(MTPC-L and MTCP-R) are positioned downstream of the magnets,
symmetrically on the left and right of the beam line. One additional
small TPC, so-called gap TPC (GTPC), is installed on the beam axis
between the vertex TPCs. The TPCs provide a measurement of charged
particle momenta $p$ with a high resolution. For the 2007 run a new
forward time of flight detector (ToF-F) was constructed in order to
extend the acceptance of the NA61/SHINE set-up for pion and kaon
identification as required for the T2K measurements \cite{nico}. The
ToF-F detector consists of 64 scintillator bars, vertically orientated, and read
out on both sides with Hamamatsu R1828 photo-multipliers. The
resolution of the ToF-F wall is $<$ 120 ps \cite{sta_rep_2008} which provides a 5 $\sigma$
$\pi$/K separation at 3 GeV/c. It is installed downstream
of the MTPC-L and MTPC-R, closing the gap between the ToF-R and ToF-L
walls. The ToF-F provides full acceptance coverage of the T2K
phase-space (parent particles generating a neutrino which
hit the far detector).

\begin{figure}[h!]
	\centering
        \includegraphics[width=.45\textwidth,height=.22\textheight]{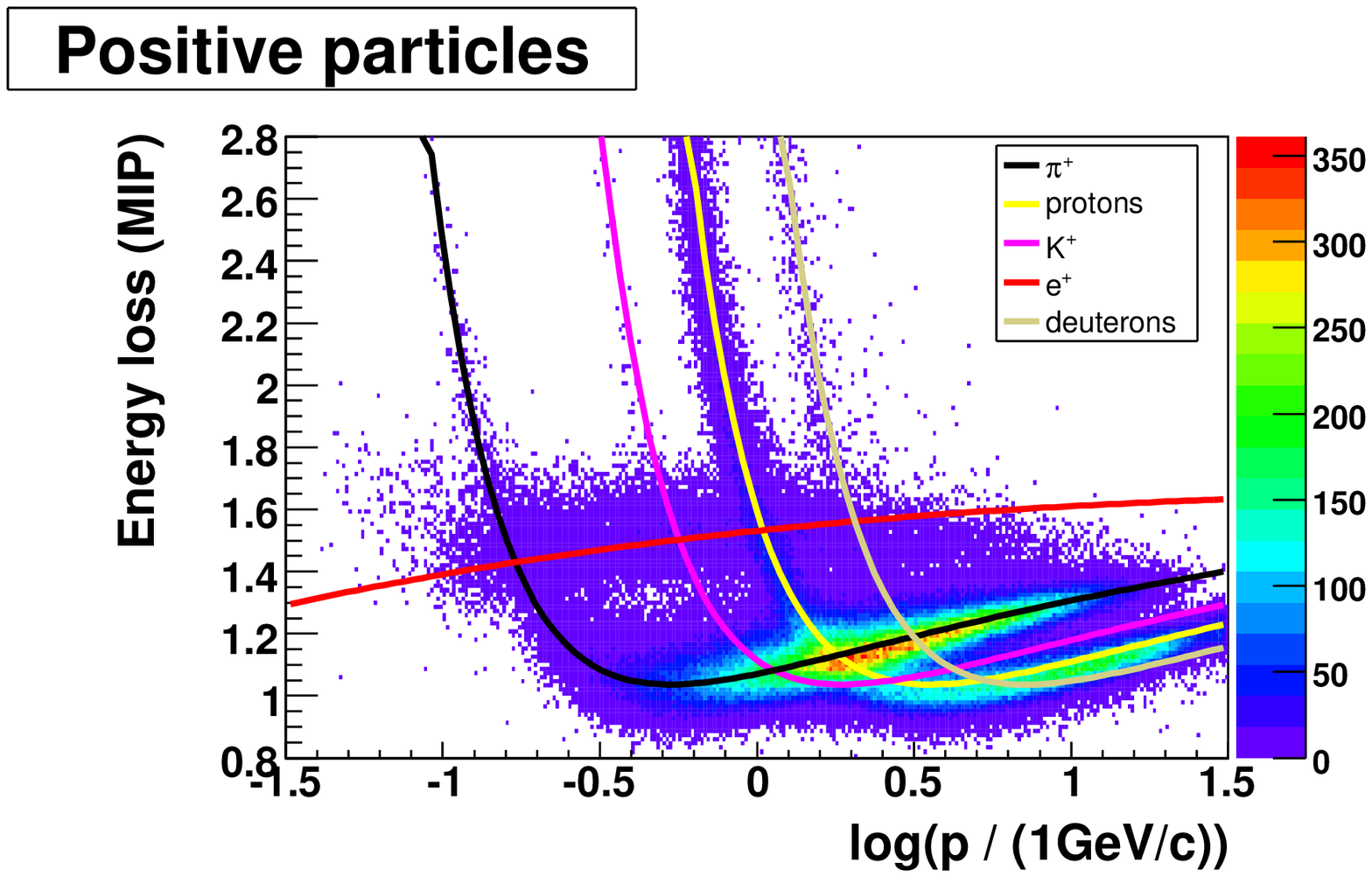}
        \includegraphics[width=.42\textwidth,height=.22\textheight]{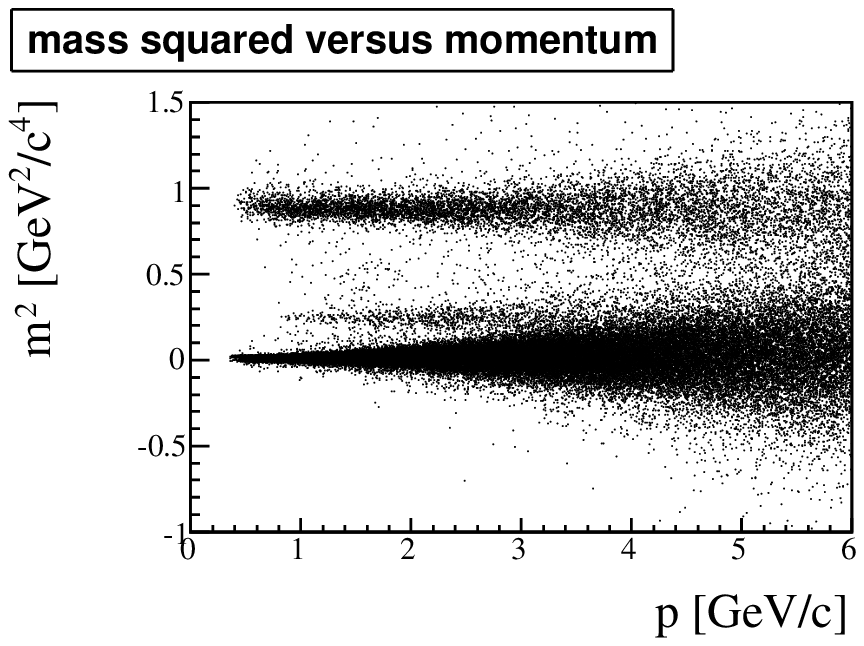}
        \includegraphics[width=.42\textwidth,height=.22\textheight]{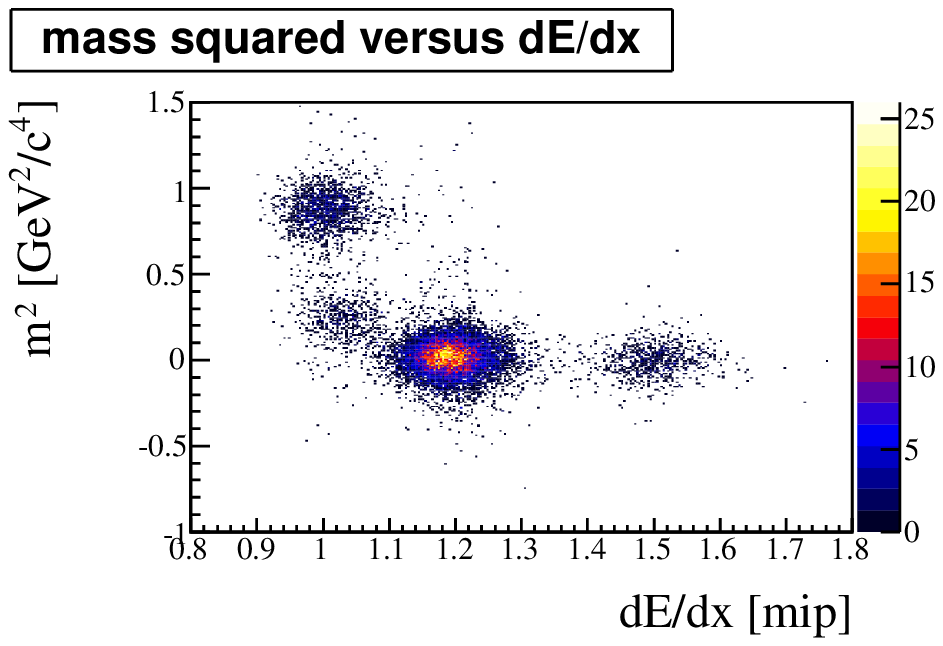}

	\caption{\label{fig:combined}[Top-Left]: dE/dx versus log(p) spectra
          and Bethe-Bloch parametization super-imposed. [Top-Right]: mass squared
          spectra from the Forward Time of flight, protons kaons and
          pions are visible. [Bottom]: mass squared versus dE/dx in
          the momentum range 3-4 GeV/c in which 4 islands
          corresponding to pions, electrons kaons and protons are
          clearly defined. High purity samples of pions can be
          extracted with this method.}
\end{figure}

%\section{Particle identification based on combined ToF + dE/dx
%  measurements}
For particles within the acceptance of the ToF detectors dE/dx
information from the TPCs is available simultaneously with the time of
flight. Fig.~\ref{fig:combined} shows dE/dx spectra for positive
particles and a mass-squared distribution from the ToF-F. 
At momenta above $\sim$4 GeV/c the separation of (e,$\pi$) from ($K$,p) is
performed essentially by dE/dx, whereas the ToF measurement is needed
to distinguish between kaons and protons. Below 4 GeV/c particle
identification can be performed almost exclusively by the ToF while
the dE/dx is needed to separate electrons. By combining both
methods pion yields can be extracted with a high purity over the whole
momentum range needed for T2K. This is demonstrated in
Fig.~\ref{fig:combined} where particles between 3 and 4
GeV/c of momentum are sorted corresponding to their dE/dx signal and the
mass squared obtained from the forward time-of-flight.

\section{Preliminary results.}
The raw pion yields are corrected step-by-step with the help of the
NA61 Geant3 based Monte-Carlo (MC). The following effects have been
accounted for: geometrical acceptance of the detector; efficiency of
the reconstruction chain; decays before reaching the ToF; ToF
detection Efficiency; pions coming from Lambda and K0s decays (called
feed-down correction); decays of pions into muons, which are
reconstructed as one track (called feed-up correction). 

In order to determine the correct number of inelastic interactions
among all triggered events some corrections have to be applied
\cite{claudia_pres}. First one has to substract the contribution of
large angle coherent elastic scatterings which pass the trigger
conditions. Secondly one needs to take into account a loss of
inelastic triggers due to a fact that secondary particle acted as
primary proton on the trigger counters. We also have to estimate the
rate of the events which take place outside of the carbon
target. 

\begin{figure}[!h]
	\centering
        \includegraphics[width=.72\textwidth,height=.36\textheight]{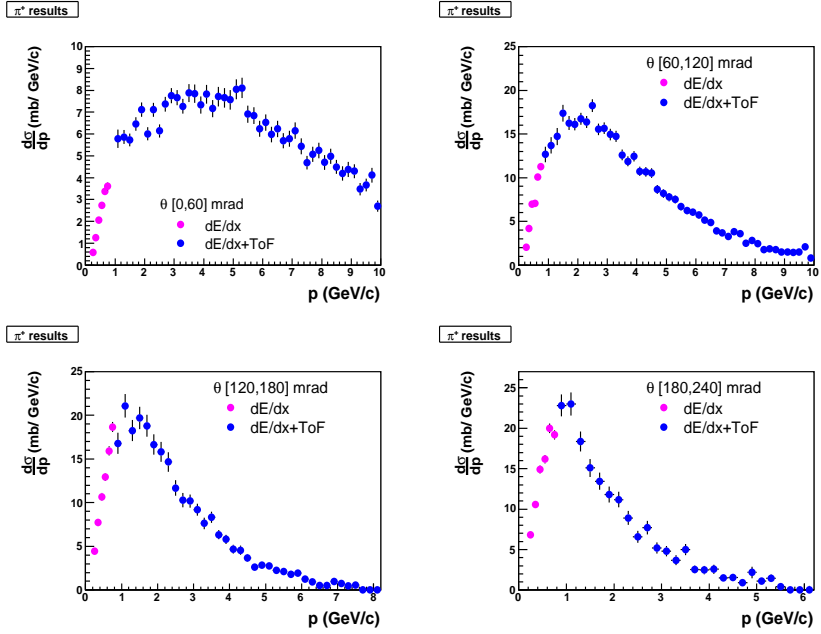}
	\caption{\label{fig:piplus_xsec}Double differential inclusive
          inelastic cross section for $\pi^{+}$ from collisions of 30 GeV protons on the thin carbon
          target. Results are shown in 4 bins of polar angle $\theta$
          between 0 and 240 mrad with statistical errors only. The results from ToF + dE/dx in blue
          are compare with the ``dE/dx only'' analysis in pink.}
\end{figure}

\begin{figure}[!h]
	\centering
        \includegraphics[width=.72\textwidth,height=.36\textheight]{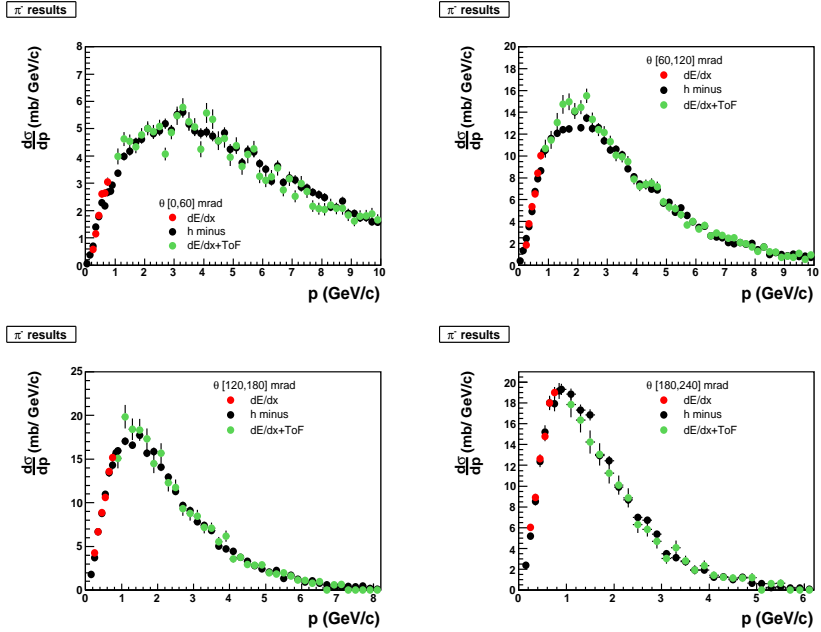}
	\caption{\label{fig:piminus_xsec}Double differential inclusive
          inelastic cross section for $\pi^{-}$. Results from 3
          analysis are shown with statistical errors only.}
\end{figure}

The double differential inclusive inelastic cross section
$\frac{d^2\sigma_{inel}}{dpd\theta}$ for $\pi^{+}$ and  $\pi^{-}$
are presented in Fig.~\ref{fig:piplus_xsec} and Fig.~\ref{fig:piminus_xsec} in
four polar angle bins. Only statistical errors are shown and for these preliminary results
systematical uncertainties are estimated to be 20\% or below.
The spectra retrieved with the combined "ToF+dE/dx" method are
compared with 2 other analysis:
\squishlist
\item in the $\pi^{+}$ case (Fig.~\ref{fig:piplus_xsec}), another analysis
  identified particles with dE/dx-only below 800 MeV/c
  \cite{maga_proc} and corrected the raw spectras with a
  global MC factor. For ToF acceptance reasons the 2 spetra do not
  overlap but the results can be checked for continuity.
\item for $\pi^{-}$ (Fig.~\ref{fig:piminus_xsec}) the results have been
  compared with a so called h-minus analysis in which all negative
  tracks were selected and yields were extracted from a global MC
  correction \cite{tomek_proc}. 
\squishend
\section{Conclusion}
Preliminary results of pion inclusive cross-section from proton carbon
interactions at 30 GeV are now available and used for the T2K flux
prediction. Results from 3 different analysis have been presented.
Work is currently in progress to reduce the systematic error of 20\%
to a level below 10\% which is required for the T2K physics. Another
much larger set of data has been collected at the end of 2009, after a
major readout upgrade, an extension of the ToF-F and a new trigger
system. This data is currently in the calibration process. With this
larger set of statistics the goal, amongst others, is to produce kaon
cross-sections results which is crucial for T2K to predict the
intrinsic $\nu_e$ contamination of the neutrino beam. Such results
would greatly improve the T2K sensitivity in its search for the last
unknown neutrino mixing angle $\theta_{13}$.

\end{document}